\documentclass[11pt]{article}
\usepackage{a4p}
\usepackage{epsfig}
\newcommand{\intLdt}  {183}
\newcommand{\intLdtfull}  {183.05}
\newcommand{\dLstat}  {0.16}
\newcommand{\dLsys}   {0.37}
\newcommand{\dLtot}   {0.40}
\newcommand{\xsecresult} {\ensuremath{16.30\pm0.34\stat\pm0.18\syst}}
\newcommand{\brqqresult} {\ensuremath{68.32\pm0.61\stat\pm0.28\syst\,\%}}
\newcommand{\rroots}  {188.635}  
\newcommand{\errroots}  {0.040}  

\newcommand{\SMDPAxs}   {16.26}
\newcommand{\dSMDPAxs}  {0.08}

\newcommand{\GENTxs}   {16.65}

\newcommand{\nsel}     {3068} 
\newcommand{\SMbrqq} {67.5}
\newcommand{\SMbrlv} {10.8}

\newcommand{\LEPII}{\mbox{LEP2}}
\newcommand{\Ebeam}{\ensuremath{E_{\mathrm{beam}}}}
\newcommand{\WW}{\ensuremath{\mathrm{W}^+\mathrm{W}^-}}
\newcommand{\epem}{\ensuremath{\mathrm{e}^+\mathrm{e}^-}}
\newcommand{\roots}{\ensuremath{\sqrt{s}}}
\newcommand{\rootsprime}{\ensuremath{\sqrt{s^\prime}}}
\newcommand{\sigccthree}{\ensuremath{\sigma_{\mathrm{WW}}}}
\newcommand{\Mw}{\ensuremath{M_{\mathrm{W}}}}
\newcommand{\Rvis}{\ensuremath{R_{\mathrm{vis}}}}
\newcommand{\stat}{\mathrm{(stat.)}}
\newcommand{\syst}{\mathrm{(syst.)}}
\newcommand{\Br}{\ensuremath{\mathrm{Br}}}
\newcommand{\Wtomn}{\mbox{$\mathrm{W}\rightarrow\mnu$}}
\newcommand{\Wtoen}{\mbox{$\mathrm{W}\rightarrow\enu$}}
\newcommand{\Wtotn}{\mbox{$\mathrm{W}\rightarrow\tnu$}}

\newcommand{\Wtoqq}{\mbox{$\mathrm{W}\rightarrow\qq$}}
\newcommand{\nue}   {\ensuremath{\nu_{e}}}
\newcommand{\numu}  {\ensuremath{\nu_{\mu}}}
\newcommand{\nutau} {\ensuremath{\nu_{\tau}}}
\newcommand{\nuell} {\ensuremath{\nu_{\ell}}}
\newcommand{\enu}   {\ensuremath{\mathrm{e}\nue}}
\newcommand{\mnu}   {\ensuremath{\mu\numu}}
\newcommand{\tnu}   {\ensuremath{\tau\nutau}}

\newcommand{\lpmnu} {\ensuremath{\ell^{\pm}\nuell}}
\newcommand{\qq}    {\ensuremath{\mathrm{q\overline{q}}}}
\newcommand{\lplm}  {\ensuremath{\ell^+\ell^-}}

\newcommand{\nunu}  {\ensuremath{\nu\overline{\nu}}}
\newcommand{\nunupri}{\ensuremath{\nu_{\ell^{\prime}}\overline{\nu}_{\ell^{\prime}}}}
\newcommand{\ffbar} {\ensuremath{f\overline{f}}}
\newcommand{\emnu}{\ensuremath{\mathrm{e^-\overline{\nu}_{e}}}}
\newcommand{\mmnu}{\ensuremath{\mu^-\overline{\nu}_{\mu}}}
\newcommand{\tmnu}{\ensuremath{\tau^-\overline{\nu}_{\tau}}}

\newcommand{\epnu}{\ensuremath{\mathrm{e^+{\nue}}}}
\newcommand{\mpnu}{\ensuremath{\mu^+{\numu}}}
\newcommand{\tpnu}{\ensuremath{\tau^+{\nutau}}}

\newcommand{\epmnu}{\ensuremath{\mathrm{e^\pm{\nue}}}}
\newcommand{\mpmnu}{\ensuremath{\mu^\pm{\numu}}}

\newcommand{\mmpnu}{\ensuremath{\mu^\mp{\numu}}}
\newcommand{\tmpnu}{\ensuremath{\tau^\mp{\nutau}}}
\newcommand{\semnu}{\ensuremath{\mathrm{e^{^-}\!\!\overline{\nu}_{e}}}}
\newcommand{\smmnu}{\ensuremath{\mu^{^-}\!\!\overline{\nu}_{\mu}}}
\newcommand{\stmnu}{\ensuremath{\tau^{^-}\!\!\overline{\nu}_{\tau}}}
\newcommand{\sepnu}{\ensuremath{\mathrm{e^{^+}\!\!{\nue}}}}
\newcommand{\smpnu}{\ensuremath{\mu^{^+}\!\!{\numu}}}
\newcommand{\stpnu}{\ensuremath{\tau^{^+}\!\!{\nutau}}}
\newcommand{\sepmnu}{\ensuremath{\mathrm{e^{^\pm}\!\!{\nue}}}}
\newcommand{\smpmnu}{\ensuremath{\mu^{^\pm}\!\!{\numu}}}

\newcommand{\smmpnu}{\ensuremath{\mu^{^\mp}\!\!{\numu}}}
\newcommand{\stmpnu}{\ensuremath{\tau^{^\mp}\!\!{\nutau}}}
\newcommand{\mpmm} {\ensuremath{{\mu}^+{\mu}^-}}
\newcommand{\tptm} {\ensuremath{{\tau}^+{\tau}^-}}
\newcommand{\epmmmp} {\ensuremath{\mathrm{e}^\pm{\mu}^\mp}}
\newcommand{\epmtmp} {\ensuremath{\mathrm{e}^\pm{\tau}^\mp}}
\newcommand{\mpmtmp} {\ensuremath{{\mu}^\pm{\tau}^\mp}}
\newcommand{\enen}{\ensuremath{\epnu\emnu}}
\newcommand{\mnmn}{\ensuremath{\mpnu\mmnu}}
\newcommand{\tntn}{\ensuremath{\tpnu\tmnu}}
\newcommand{\enmn}{\ensuremath{\epmnu\mmpnu}}
\newcommand{\entn}{\ensuremath{\epmnu\tmpnu}}
\newcommand{\mntn}{\ensuremath{\mpmnu\tmpnu}}
\newcommand{\enens}{\ensuremath{\enu\enu}}
\newcommand{\mnmns}{\ensuremath{\mnu\mnu}}
\newcommand{\tntns}{\ensuremath{\tnu\tnu}}
\newcommand{\enmns}{\ensuremath{\enu\mnu}}
\newcommand{\entns}{\ensuremath{\enu\tnu}}
\newcommand{\mntns}{\ensuremath{\mnu\tnu}}
\newcommand{\senen}{\ensuremath{\sepnu\semnu}}
\newcommand{\smnmn}{\ensuremath{\smpnu\smmnu}}
\newcommand{\stntn}{\ensuremath{\stpnu\stmnu}}
\newcommand{\senmn}{\ensuremath{\sepmnu\smmpnu}}
\newcommand{\sentn}{\ensuremath{\sepmnu\stmpnu}}
\newcommand{\smntn}{\ensuremath{\smpmnu\stmpnu}}
\newcommand{\lnln}{\ensuremath{\ell^+{\nu}_{\ell}\ell^{\prime -}\overline{\nu}_{\ell^\prime}}}
\newcommand{\qqen}{\ensuremath{\qq\enu}}
\newcommand{\qqmn}{\ensuremath{\qq\mnu}}
\newcommand{\qqtn}{\ensuremath{\qq\tnu}}
\newcommand{\qqln}{\ensuremath{\qq\lpmnu}}
\newcommand{\qqqq}{\ensuremath{\qq\qq}}
\newcommand{\lnlns}{\ensuremath{\ell\nu\ell\nu}}
\newcommand{\qqlns}{\ensuremath{\mathrm{qq}\ell\nu}}
\newcommand{\qqqqs}{\ensuremath{\mathrm{qqqq}}}

\newcommand{\WWqqen}{\mbox{\WW$\rightarrow$ \qqen}}
\newcommand{\WWqqmn}{\mbox{\WW$\rightarrow$ \qqmn}}
\newcommand{\WWqqtn}{\mbox{\WW$\rightarrow$ \qqtn}}
\newcommand{\WWlnln}{\mbox{\WW$\rightarrow$ \lnln}}
\newcommand{\WWqqln}{\mbox{\WW$\rightarrow$ \qqln}}
\newcommand{\WWqqqq}{\mbox{\WW$\rightarrow$ \qqqq}}
\newcommand{\llnn}{\ensuremath{\lplm\nunupri}}

\newcommand{\qqnn}{\ensuremath{\qq\nunu}}
\newcommand{\qqll}{\ensuremath{\qq\lplm}}
\newcommand{\qqee}{\ensuremath{\qq\epem}}
\newcommand{\qqmm}{\ensuremath{\qq\mpmm}}
\newcommand{\ZZqqqq}{\ensuremath{\mathrm{ZZ}\rightarrow\qq\qq}}
\newcommand{\eell}{\ensuremath{\epem\lplm}}
\newcommand{\eeff}{\ensuremath{\epem\ffbar}}
\newcommand{\eemm}{\ensuremath{\epem\mu^+\mu^-}}
\newcommand{\eetoqq}{\ensuremath{\epem\rightarrow\qq}}
\newcommand{\eetoee}{\ensuremath{\epem\rightarrow\epem}}
\newcommand{\eetomm}{\ensuremath{\epem\rightarrow\mpmm}}
\newcommand{\eetott}{\ensuremath{\epem\rightarrow\tptm}}
\newcommand{\eetonunuG}{\ensuremath{\epem\rightarrow\nunu\gamma(\gamma)}}
\newcommand{\Zz}       {\ensuremath{{\mathrm{Z}^0}}}

\newcommand{\nngg}     {\ensuremath{\nunu\gamma(\gamma)}}

\newcommand{\Zll}      {\ensuremath{\Zz\rightarrow\lplm}}
\newcommand{\Zqq}      {\ensuremath{\Zz\rightarrow\qq}}
\newcommand{\Zgamma}   {\ensuremath{\Zz/\gamma}}

\newcommand{\Wenu}     {\ensuremath{\mathrm{W}\enu}}
\newcommand{\Zee}      {\ensuremath{\mathrm{Z}\epem}}
\newcommand{\Znunu}    {\ensuremath{\mathrm{Z}\nunu}}
\newcommand{\qqgg}     {\ensuremath{\qq \mathrm{gg}}}

\newcommand{\CC}{\mbox{{\sc CC03}}}

\newcommand{\JETSET}{\mbox{J{\sc etset}}}
\newcommand{\KORALW}{\mbox{K{\sc oralw}}}
\newcommand{\KORALZ}{\mbox{K{\sc oralz}}}
\newcommand{\EXCALIBUR}{\mbox{E{\sc xcalibur}}}
\newcommand{\GRC}{\mbox{grc4f}}
\newcommand{\PYTHIA}{\mbox{P{\sc ythia}}}

\newcommand{\GENTLE}{\mbox{G{\sc entle}}}
\newcommand{\BHWIDE}{\mbox{B{\sc hwide}}}
\newcommand{\PHOJET}{\mbox{P{\sc hojet}}}
\newcommand{\HERWIG}{\mbox{H{\sc erwig}}}
\newcommand{\KK}{\mbox{KK2f}}
\newcommand{\YFSWW}{\mbox{Y{\sc fs}WW}}
\newcommand{\RACWW}{\mbox{R{\sc acoon}WW}}
\newcommand{\Pt}{\ensuremath{p_{\mathrm{T}}}}
\newcommand{\Xt}{\ensuremath{x_{\mathrm{T}}}}
\newcommand{\WQCD}{\ensuremath{W_{420}}}
\newcommand{\WCC}{\ensuremath{W_{\mathrm{CC03}}}}
\newcommand{\Vij} {\mbox{$|\mathrm{V}_{ij}|$}}
\newcommand{\Vud} {\mbox{$|\mathrm{V}_{\mathrm{ud}}|$}}
\newcommand{\Vus} {\mbox{$|\mathrm{V}_{\mathrm{us}}|$}}
\newcommand{\Vcd} {\mbox{$|\mathrm{V}_{\mathrm{cd}}|$}}
\newcommand{\Vcb} {\mbox{$|\mathrm{V}_{\mathrm{cb}}|$}}
\newcommand{\Vub} {\mbox{$|\mathrm{V}_{\mathrm{ub}}|$}}
\newcommand{\Vcs} {\mbox{$|\mathrm{V}_{\mathrm{cs}}|$}}
\newcommand{\sig} {\ensuremath{\mathrm{sig}}}
\newcommand{\bgd} {\ensuremath{\mathrm{bgd}}}
\newcommand{\esig} {\ensuremath{\varepsilon^{\sig}}}
\newcommand{\sbgd} {\ensuremath{\sigma^{\bgd}}}
\newcommand{\etal}    {\mbox{{\it et al.}}}
\newcommand{\Journal}[4] {{#1} \textbf{#2} (#3) {#4}}
\newcommand{\PL}  {Phys. Lett. }
\newcommand{\ZP}  {Z. Phys.}
\newcommand{\EPJ} {Eur. Phys. J.}
\newcommand{\NIM} {Nucl. Instr. and Meth.}

\newcommand{\PRL} {Phys. Rev. Lett.}
\newcommand{\PR}  {Phys. Rev.}
\newcommand{\NP}  {Nucl. Phys.}
\newcommand{\CPC} {Comp. Phys. Comm.}
\newcommand{\CIP} {Comp. in Phys.}
\def\opalabbiendi{OPAL Collaboration, G.\ Abbiendi \etal}
\def\opalackerstaff{OPAL Collaboration, K.\ Ackerstaff \etal}
\def\opalalexander{OPAL Collaboration, G.\ Alexander \etal}

\def\opalakrawy{OPAL Collaboration, M.Z.\ Akrawy \etal}
\def\opalahmet{OPAL Collaboration, K.\ Ahmet \etal}
\def\opalanderson{OPAL Collaboration, S.\ Anderson \etal}
\begin{document}
\begin{titlepage}
\begin{center}
  {\Large   EUROPEAN ORGANIZATION FOR NUCLEAR RESEARCH}
\end{center}
\bigskip
\begin{flushright}
  CERN-EP-2000-101 \\
  20 July 2000
\end{flushright}
\bigskip\bigskip\bigskip\bigskip\bigskip
\begin{center}
  \addtolength{\baselineskip}{4mm}
  {\huge\bf\boldmath \WW\ Production Cross Section}\\
  {\huge\bf\boldmath and W Branching Fractions}\\
  {\huge\bf\boldmath in \epem\ Collisions at 189~GeV}
\end{center}
\bigskip\bigskip
\begin{center}
{\LARGE The OPAL Collaboration}
\end{center}
\bigskip\bigskip\bigskip
\begin{center}{\large\bf Abstract}\end{center} 
From a data sample of \intLdt~pb$^{-1}$ recorded at a center-of-mass energy 
of $\roots = 189$~GeV with the OPAL detector at LEP, \nsel\ W-pair 
candidate events are selected.
Assuming Standard Model W boson decay branching fractions,
the W-pair production cross section is measured to be
\sigccthree=\xsecresult~pb. 
When combined with previous OPAL measurements, the W boson branching 
fraction to hadrons is determined to be \brqqresult\ assuming 
lepton universality.
These results are consistent with Standard Model expectations.
\bigskip\bigskip\bigskip\bigskip
\bigskip\bigskip
\begin{center}
  {\large (Submitted to Physics Letters B)}
\end{center}
\end{titlepage}
%
%
\begin{center}{\Large The OPAL Collaboration
}\end{center}
\begin{center}{
G.\thinspace Abbiendi$^{  2}$,
K.\thinspace Ackerstaff$^{  8}$,
C.\thinspace Ainsley$^{  5}$,
P.F.\thinspace {\AA}kesson$^{  3}$,
G.\thinspace Alexander$^{ 22}$,
J.\thinspace Allison$^{ 16}$,
K.J.\thinspace Anderson$^{  9}$,
S.\thinspace Arcelli$^{ 17}$,
S.\thinspace Asai$^{ 23}$,
S.F.\thinspace Ashby$^{  1}$
% ...
}\end{center}\bigskip
\bigskip
$^{  1}$School of Physics and Astronomy, University of Birmingham,
Birmingham B15 2TT, UK
\newline
$^{  2}$Dipartimento di Fisica dell' Universit\`a di Bologna and INFN,
I-40126 Bologna, Italy
\newline
$^{  3}$Physikalisches Institut, Universit\"at Bonn,
D-53115 Bonn, Germany
\newline
% ...
$^{  5}$Cavendish Laboratory, Cambridge CB3 0HE, UK
\newline
% ...
$^{  8}$CERN, European Organisation for Nuclear Research,
CH-1211 Geneva 23, Switzerland
\newline
$^{  9}$Enrico Fermi Institute and Department of Physics,
University of Chicago, Chicago IL 60637, USA
\newline
% ...
$^{ 16}$Department of Physics, Schuster Laboratory, The University,
Manchester M13 9PL, UK
\newline
$^{ 17}$Department of Physics, University of Maryland,
College Park, MD 20742, USA
\newline
% ...
$^{ 22}$Department of Physics and Astronomy, Tel Aviv University,
Tel Aviv 69978, Israel
\newline
$^{ 23}$International Centre for Elementary Particle Physics and
Department of Physics, University of Tokyo, Tokyo 113-0033, and
Kobe University, Kobe 657-8501, Japan
\newline
% ...
\newpage
%
%
\section{Introduction}

In 1996, the LEP collider at CERN entered a new phase of operation,
\LEPII, with the first \epem\ collisions above the \WW\ production
threshold at $\roots = 161$~GeV.
By 1998, with the installation of additional super-conducting 
radio-frequency accelerating cavities, the center-of-mass collision 
energy of the LEP collider was increased to $\roots = 189$~GeV.
This paper describes the measurement of the \WW\ production 
cross section and the W boson branching fractions using 
\intLdt~pb$^{-1}$ of data recorded by the OPAL detector during 
the 1998 LEP run.
This measurement provides an important test of the non-Abelian
nature of the electroweak interaction, as the \WW\ production 
cross section above threshold is sensitive to the couplings 
between the weak gauge bosons.
In addition, with the large sample of W bosons produced in 
1998, more precise tests of the weak charged-current interaction 
can be made in the measurement of the W boson branching
fractions to leptons and hadrons.

In the Standard Model, \WW\ events are expected to decay into fully
leptonic (\WW$\rightarrow$ \lnln), semi-leptonic (\WWqqln), 
or fully hadronic (\WWqqqq) final states with predicted 
branching fractions of 10.6\%, 43.9\%, and 45.6\% 
respectively~\cite{bib:LEP2YR}.
Three separate selections, described in  
Sections~\ref{sec:lnln}--\ref{sec:qqqq}, are used in this analysis
to identify candidate \WW\ events by their final state topologies.
For the \lnlns\ and \qqlns\ event selections, events are further
classified according to charged lepton type.
In total, \WW\ candidate events are exclusively selected in 
one of ten possible final states $(6\times\lnlns$, $3\times\qqlns$,
and $1\times\qqqqs)$.

From the observed event rates in these ten channels, measurements
of the W boson branching fractions and the total \WW\ production
cross section are performed as described in Section~\ref{sec:results}.
The branching fraction measurements at $\roots = 189$~GeV are also 
combined with previous OPAL results from data collected at 
$\roots = 161$~GeV~\cite{bib:opalmw161},
$\roots = 172$~GeV~\cite{bib:opalmw172}, and 
$\roots = 183$~GeV~\cite{bib:opaltgc183}.
%
%
\section{Data and Monte Carlo Models}

The OPAL detector has been described in detail in previous 
publications~\cite{bib:opaldet}.
The data reconstruction, luminosity measurement, Monte Carlo models,
and detector simulation used for this analysis
are identical to those used in previous OPAL \WW\ cross-section
measurements~\cite{bib:opalmw172,bib:opaltgc183}.
The accepted integrated luminosity, evaluated using small angle Bhabha
scattering events observed in forward calorimeters, 
is $\intLdtfull\pm\dLstat \mathrm{(stat.)} 
\pm\dLsys \mathrm{(syst.)}$~pb$^{-1}$~\cite{bib:lumi183}.
The mean center-of-mass energy for the data sample is 
$\roots = \rroots \pm \errroots$~GeV~\cite{bib:energy189}.

The semi-analytic program 
\GENTLE~2.0~\cite{bib:GENTLE} has been used to 
calculate the \WW\ cross section
$\sigccthree = \GENTxs$~pb at $\roots = \rroots$~GeV assuming
a W boson mass of $\Mw=80.41$~GeV~\cite{bib:pdg}.
The estimated theoretical uncertainty on the \GENTLE\ prediction
is 2\%.
Recent theoretical calculations which include a more complete
treatment of ${\cal O}(\alpha)$ radiative corrections through the 
double pole approximation~\cite{bib:DPA} are now available in 
the \YFSWW~\cite{bib:YFSWW} and \RACWW~\cite{bib:RAC} Monte Carlo 
generators.
These new calculations predict a \WW\ cross section of 16.27~pb
and 16.25~pb respectively, with a reduced theoretical 
uncertainty of 0.42\%.
Even though \YFSWW\ and \RACWW\ differ in their implementation
of ${\cal O}(\alpha)$ radiative corrections, these two calculations 
are in agreement at the level of 0.1\%, and predict a rate
which is (2.3--2.4)\% lower than the older \GENTLE\ estimate. 

A \WW\ cross section of $\SMDPAxs\pm\dSMDPAxs$~pb is used 
throughout this paper to determine the expected number of 
\WW\ events predicted by the Standard Model.
This value is chosen to be representative of these improved
calculations, while covering the expected range of theoretical 
uncertainty quoted by either \YFSWW\ or \RACWW.

A number of additional Monte Carlo generators are used in this analysis
to provide estimates of signal efficiencies and 
expected backgrounds from other Standard Model processes.
Unless otherwise noted, all Monte Carlo event samples have been processed 
through a detailed simulation of the OPAL detector~\cite{bib:GOPAL}.

\begin{figure}
  \begin{center}
    \epsfig{file=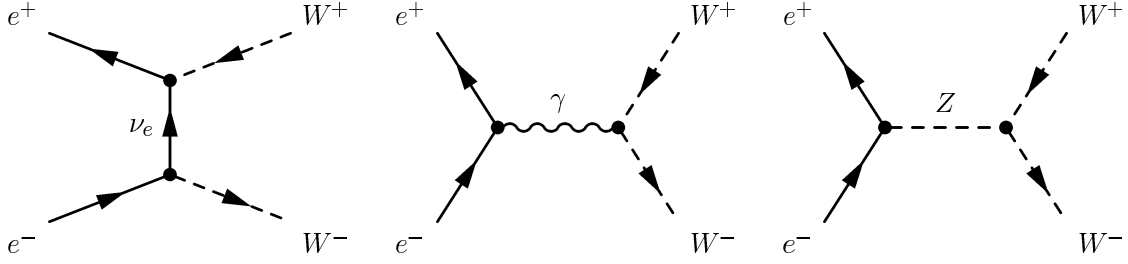}
    \caption{
      The \CC\ diagrams for W-pair production.
      \label{fig:CC03}
      }
  \end{center}
\end{figure}

In this paper, \WW\ events are defined in terms of the \CC\ class of 
production diagrams shown in Figure~\ref{fig:CC03} following the 
notation of~\cite{bib:LEP2YR}.
These amplitudes, namely the $t$-channel \nue\ exchange and 
$s$-channel \Zgamma\ exchange, provide a natural definition of resonant
W-pair production, even though other non-\CC\ diagrams contribute to the
same four-fermion final states.
The efficiency for selecting \CC\ signal events is estimated using
the \KORALW~1.42~\cite{bib:KORALW} Monte Carlo generator, 
with the predictions of \EXCALIBUR~\cite{bib:EXCALIBUR}, 
\PYTHIA~\cite{bib:PYTHIA}, and \HERWIG~\cite{bib:HERWIG} being used to 
assess possible systematic uncertainties on the theoretical prediction.

To extract the \CC\ cross section from the data, the expected difference 
between the complete four-fermion production rate and the rate predicted
using only the \CC\ diagrams is treated as a background and 
subtracted from the observed cross section.
This four-fermion background is estimated using the 
\KORALW, \EXCALIBUR, and \GRC~\cite{bib:GRC4F} four-fermion
Monte Carlo generators,\footnote{
The \KORALW\ four-fermion generator uses the \GRC\ four-fermion
matrix elements, but other details of the event generation differ.
}
and includes contributions from both non-\CC\ four-fermion
final states and interference between the \CC\ and non-\CC\ amplitudes.
For the \lnlns\ final states, where this additional non-\CC\ four-fermion 
contribution is most pronounced, inclusive four-fermion cross sections 
are also quoted following a signal definition described in 
Section~\ref{sec:lnln4f}.

Additional backgrounds from two-fermion final states are estimated using 
the \PYTHIA, \HERWIG, and \KK~\cite{bib:KK2F} Monte Carlo generators for 
the \eetoqq\ process, 
\KORALZ~\cite{bib:KORALZ} for the \eetomm, \eetott, and \eetonunuG\ 
processes, and \BHWIDE~\cite{bib:BHWIDE} for the \eetoee\ process.
Backgrounds from two-photon interactions are evaluated using \PYTHIA, 
\HERWIG, \PHOJET~\cite{bib:PHOJET}, and the Vermaseren
program~\cite{bib:VERMASEREN}.
%
%
\section{\boldmath \WWlnln\ Event Selection}
\label{sec:lnln}

Fully leptonic \WW\ events are identified as a pair of
charged leptons with significant missing transverse momentum.
In previous OPAL results~\cite{bib:opalmw172,bib:opaltgc183}, events
were considered as \WWlnln\ candidates if they were selected by 
either of two independent selection algorithms.
For the results presented in this paper, however, an improved 
version of the OPAL acoplanar di-lepton selection II presented 
in references~\cite{bib:ACOPLL,bib:ACOPLL189} is used exclusively.
The new selection has an efficiency of 82\% for \WW\ events at 
$\roots = 189$~GeV (was approx 72\% for selection II in \cite{bib:ACOPLL}) 
and a background of 38~fb (was 66~fb in \cite{bib:ACOPLL189}).
The main improvements are as follows:
\begin{itemize}
  \item simplifying the kinematic cuts such that they are lepton flavor
        independent and based primarily on requirements of significant 
        missing transverse momentum ($\Pt$);
  \item using the recently installed forward scintillating tile 
        counters~\cite{bib:tiles} to efficiently reject backgrounds from 
        otherwise undetected forward muons;
  \item requiring that the measured missing $\Pt$\ could 
        not easily be faked by mis-measurements of the track momenta.
\end{itemize}
Since the characteristic \lnlns\ topology is shared by other 
non-\CC\ Standard Model processes as well as many manifestations of new
physics beyond the Standard Model, no attempt has been made
in this selection to discriminate \CC\ production from other
sources.
Rather, the selection is developed to be globally efficient for 
any mechanism which produces two charged leptons and missing 
transverse momentum
in the final state while rejecting the main backgrounds from
two-fermion production and two-photon interactions.

To be selected as \WWlnln\ candidates, events must pass a series
of cuts designed to isolate the signal events while rejecting the
dominant backgrounds.
After rejecting all high multiplicity events, ``jets'' are found in
each candidate event using a cone algorithm applied to the observed
tracks and calorimeter clusters.
A cone half-opening angle 
of 20 degrees and a jet energy threshold of 2.5~GeV is used.
Only those events with between one and three identified jets are 
considered further, and a different series of additional cuts is
applied depending upon the number of jets observed.

In the most common di-jet case, where the charged lepton candidates
are cleanly separated from each other, the most important cuts require
a minimum acollinearity angle between the two leptons 
$(\theta_{\mathrm{acol}} > 5^{\circ})$, and a minimum scaled transverse 
momentum $(\Xt = \Pt/\Ebeam > 5\%)$.
Many additional cuts are applied to reject specific background
processes, including the rejection of events with significant
activity in the forward scintillating tile counters consistent 
with an otherwise undetected forward muon from \eemm\ production.

The tri-jet selection is designed to retain efficiency for $\lnlns\gamma$
final states where the photon forms a third jet, with
additional cuts applied to reject two-fermion $\tau^+\tau^-\gamma$
production.
The mono-jet selection is designed to select additional events where
both leptons are reconstructed within the same cone, or where one 
lepton is only partially reconstructed in the forward direction.
Tighter cuts are required on \Xt\ to suppress backgrounds
from two-photon interactions, and event timing cuts are used to 
eliminate backgrounds from cosmic rays entering the detector.

The classification of the selected \lnlns\ events into 
di-lepton class is initially based on the observed lepton 
properties, as well as the observed track multiplicity in each jet.
This classification is further refined by momentum cuts
dependent upon the acollinearity angle
such that identified $e$ or $\mu$ leptons with low momentum 
consistent with $\tau\rightarrow\ell\nuell\nutau$ decays are 
reclassified as $\tau$ lepton candidates.
These cuts are effective due to the correlation between lepton
energy and decay angle from the parent W boson.

The inclusive \lnlns\ selection is estimated to be 
$(82.1 \pm 1.2)\%$ efficient for \WWlnln\ events, where the error 
indicates the systematic uncertainty.
The detailed efficiency matrix listing the selection efficiency 
of each di-lepton selection class for a specific
\WWlnln\ event type is shown in Table~\ref{tab:eff}.
A correction of $(-0.9 \pm 0.4)\%$ has been applied to the efficiency
predicted by \KORALW\ Monte Carlo samples to account 
for unmodelled beam-related backgrounds in the forward detectors.
Since significant activity in the forward detectors is used as a veto
against various background classes, like the rejection of 
\eemm\ mentioned above, this beam-related detector occupancy 
causes a reduction of the selection efficiency which
is estimated from randomly triggered beam crossings.
A variety of other possible systematic effects have been considered
including the dependence on W mass, beam energy, trigger efficiency,
and several aspects of the 
detector modeling in the Monte Carlo simulation.
All are found to be small ($< 0.4\%$ each), and a 
total relative uncertainty on the selection efficiency of $1.5\%$ is 
assessed.
This systematic uncertainty is small compared to the expected 
statistical errors.

Backgrounds to the \lnlns\ event selection can be grouped
into three distinct classes.
The first class consists of backgrounds from processes which 
do not contain two leptons and two neutrinos in the final state.
Predominantly $\tau$-pair and two-photon (\eell) production, 
this background class contributes an expected $38\pm10$~fb to 
the total selected \lnlns\ rate.
The second class consists of an irreducible background from 
\llnn\ final states which can only be produced by neutral 
current diagrams as the neutrinos are of a different lepton
species than the charged leptons.
Since the neutrinos are unobserved, these final states are 
indistinguishable from the signal events in terms of the event
topology.
This class contributes an additional $45\pm2$~fb of background 
to the inclusive \lnlns\ selection.
The final class of background is the difference between the 
complete four-fermion cross section and the theoretically 
predicted rate from \CC\ diagrams alone for \lnlns\ final states.
This includes neutral-current processes in the final states where
the two charged leptons are of the same type 
$(\ell^{+}\nu_{\ell}\ell^{-}\overline{\nu}_{\ell})$
and other four-fermion processes when there is an electron
in the final state $(e^{\pm}\nu_{e}\ell^{\mp}\nu_{\ell})$.
This non-\CC\ production contributes a large cross section of
$77\pm10$~fb which is treated as a background
in the \CC\ selection and is also largely irreducible
within the detector acceptance.
The errors on the accepted background rates include all
systematic uncertainties, including the effects of limited
Monte Carlo statistics.
A detailed breakdown of the accepted background cross-sections
for the six di-lepton classes identified in the \lnlns\ selection
is shown in Table~\ref{tab:backlnln}.

The dominant systematic uncertainty associated with the \lnlns\ 
background estimate is due to the four-fermion correction.
The accepted four-fermion background rate is estimated from the
difference observed in \KORALW\ four-fermion and \CC\ samples with 
equivalent \EXCALIBUR\ samples providing a cross-check.
The accepted background rates from all processes for the \lnlns\ event 
selection are shown in Tables~\ref{tab:backlnln} and~\ref{tab:back}.

\begin{figure}
  \begin{center}
    \epsfig{file=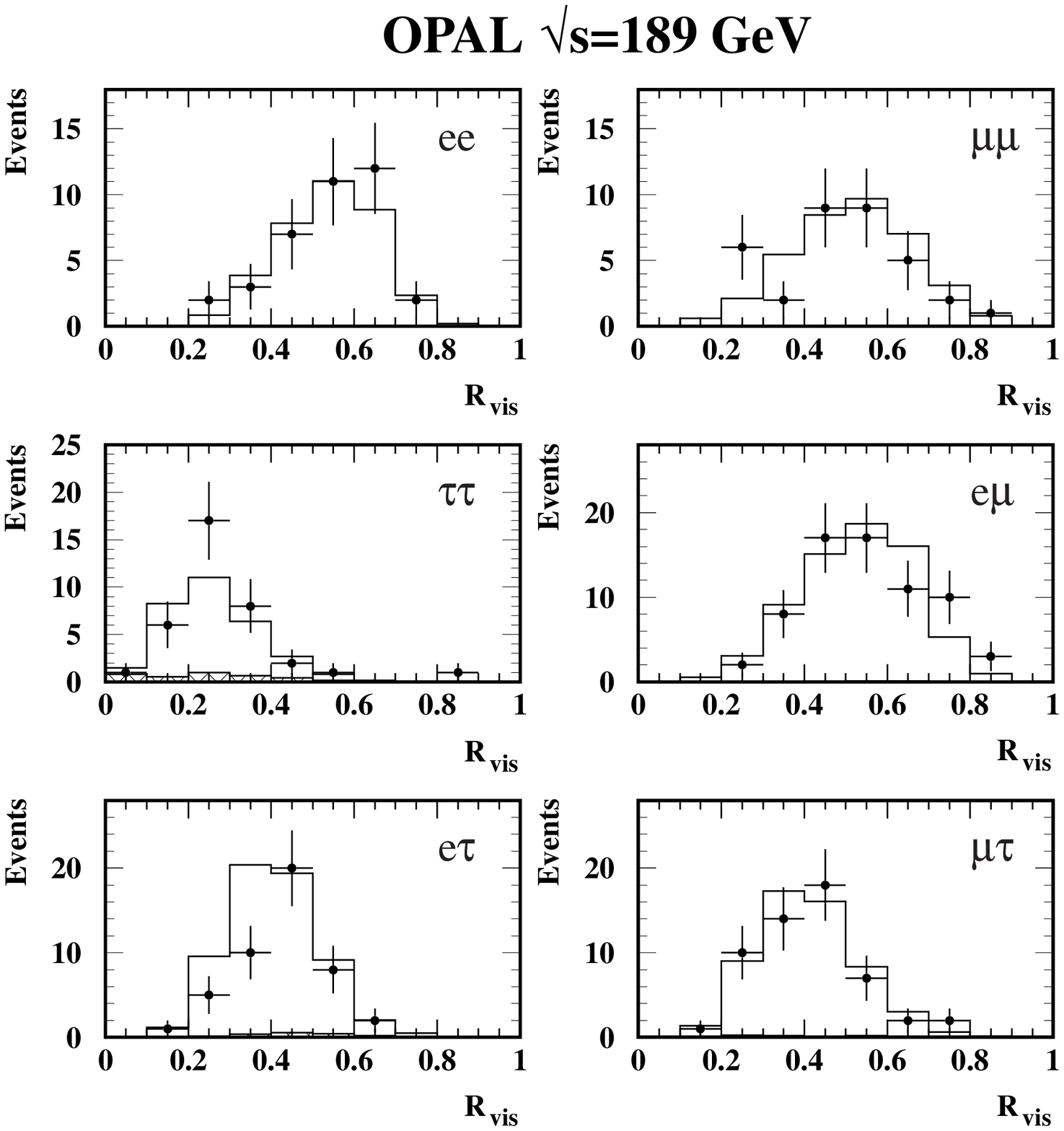, width=\textwidth}
    \caption{ 
      Distributions of visible energy scaled to the center-of-mass 
      energy, \Rvis, for the six di-lepton classes selected as \WWlnln. 
      The data are shown as the points with statistical error bars. 
      The total Monte Carlo expectation is shown as the histogram 
      with the non-\lplm\nunu\ background contribution shown by the
      hatched histogram.
      \label{fig:lnlnsel}
      }
  \end{center}
\end{figure}

A total of 276 events are selected in the data, with
$29 \pm 3$  expected from all background classes.
Figure~\ref{fig:lnlnsel} shows distributions of the reconstructed 
visible energy fraction for the six individual di-lepton 
classifications.
%
%
\section{\boldmath \WWqqln\ Event Selection}
\label{sec:qqln}

The \WWqqln\ selection consists of three separate selections, one
for each type of semi-leptonic decay. 
Only those events which are not already selected as \WWlnln\ candidates
are considered by these selections, and the \WWqqtn\ selection 
is only applied to those events which fail both the \WWqqen\ 
and \WWqqmn\ selections.

The \WWqqln\ event selection for the 189~GeV data is based on
that described in detail in previous 
publications~\cite{bib:opalmw172,bib:opaltgc183}.
The selection consists of five stages, which can be summarized as
\begin{itemize}
   \item a loose preselection to remove events with low multiplicity or
         little visible energy;
   \item identification of the observed track in the event most
         consistent with being the leptonic decay of a W boson; 
   \item separate likelihood selections for \WWqqen, \WWqqmn, and \WWqqtn;
   \item re-classification of \qqtn\ events which are identified 
         by the \qqen\ and \qqmn\ selections;
   \item rejection of four-fermion backgrounds. 
\end{itemize}
The first three stages are optimized for the rejection of the 
\eetoqq\ background which has an expected cross section about six
times larger than the W-pair production cross section at 189~GeV. 
The \WWqqln\ likelihood selections have a significant efficiency 
for other four-fermion processes, {\em e.g.} \qqen\ final states
produced by the single W (\Wenu) diagrams and \qqll production.
For this reason additional cuts are applied to events passing the 
likelihood selections to reduce backgrounds from these processes.  

The four-fermion background rejection consists of three separate parts. 
Firstly, cuts are applied to selected \WWqqen\ and \WWqqmn\ events to
reduce backgrounds from \qqee\ and \qqmm\ final states 
where both leptons are observed in the detector.  
Secondly, the \WWqqtn\ selection accepts
approximately 40\% of hadronically decaying single W events 
(\qqen) where the electron is produced in the
far forward region  beyond the experimental acceptance. 
In these events a fragmentation track is mis-identified 
as a $\tau$ lepton decay product. 
To reduce this background, an additional likelihood
selection is applied which separates \WWqqtn\ from \Wenu. 
Finally, background in the \WWqqen\ selection from the \Zee\ final state, 
where the \Zz\ decays hadronically and one electron is far forward, 
is reduced with two kinematic fits, the first using the hypothesis that 
the event is \WWqqen\ and the second using the \Zee\ hypothesis.

In addition to the likelihood selections, 
cut based selections are used to identify \WWqqen\ and \WWqqmn\
events where the lepton track is either poorly reconstructed or 
is beyond the tracking acceptance. 
These `trackless' selections require clear evidence of an electron or muon
in the calorimeter or muon chambers consistent with the kinematics of a
\WWqqln\ event, without explicitly demanding a reconstructed track.
These additional selections improve the overall efficiency by approximately
3\% (5\%) for \WWqqen\ (\WWqqmn) events, while reducing the 
systematic uncertainties associated with the modeling of 
the forward tracking acceptance.

The inclusive \qqlns\ selection is estimated to be 
$(86.8 \pm 0.9)\%$ efficient for \WWqqln\ events, as
predicted by \KORALW\ Monte Carlo samples.
The efficiencies of the \WWqqln\ selection for the individual 
channels are given in Table~\ref{tab:eff}.
These efficiencies include small corrections ($0.5\%$) which 
account for observed differences between data 
and the Monte Carlo simulation.
These corrections are obtained using `mixed events' formed by 
superimposing \Zqq\ multihadronic events and hemispheres from 
\Zll\ lepton pairs recorded at $\roots = 91$~GeV as 
described previously~\cite{bib:opalmw172}.
Small corrections ($0.3\%$) are also applied to account for
tracking losses which are not modeled by
the Monte Carlo simulation of the OPAL detector.
These corrections are determined by studying 
$\Zz\rightarrow\lplm$ events.
The effect of detector occupancy from beam-related backgrounds
has also been evaluated.
 
Possible biases due to hadronization uncertainties are studied with
fully simulated Monte Carlo \WWqqln\ samples where the hadronization
process is modeled using either \JETSET\ or \HERWIG.
Other systematics are evaluated by comparing samples generated 
with different Monte Carlo generators
(\KORALW, \PYTHIA, \EXCALIBUR, and \GRC).
In each case, the largest observed difference between generators
is taken as an estimate of the systematic uncertainty.
Table~\ref{tab:effsys} lists the various contributions to the 
systematic uncertainty on the  selection efficiency.

Table~\ref{tab:back} shows the background cross sections and
total uncertainties for the three \qqlns\ selections.  
The systematic errors on the expected background cross sections
are dominated by differences between data and Monte Carlo for the 
two-fermion backgrounds and by differences between
generators in the case of the four-fermion backgrounds. 
The systematic errors on the four-fermion backgrounds were estimated 
by comparing the expectations of \KORALW\ and \EXCALIBUR.

The dominant background in the \qqlns\ selection is from di-jet 
production, predominantly the \eetoqq\ and single W processes, 
where a particle produced during hadronization is incorrectly 
identified as a prompt lepton.
The Monte Carlo estimate of this background rate is checked using 
control samples constructed from the data directly.
For the \eetoqq\ background, `fake' events are constructed by boosting
hadronic \Zz\ events recorded at $\roots = 91$~GeV to the \rootsprime\ 
distribution expected of quark pairs at $\roots = 189$~GeV.
This boost procedure is applied to both \Zz\ data and \Zz\ Monte Carlo
samples, with the ratio of selected events in each \qqlns\ channel
being used to assign a systematic uncertainty of 15\%.
For the \Wenu\ and \qqnn\ backgrounds, which are large in 
the \qqtn\ channel, a control sample is constructed from selected 
\qqln\ events by discarding the selected lepton track.
Again, the observed ratio in selected events between data and
Monte Carlo samples is used to assign a systematic uncertainty
of 10\% to this background source.
  
\begin{figure}
 \begin{center}
   \epsfig{file=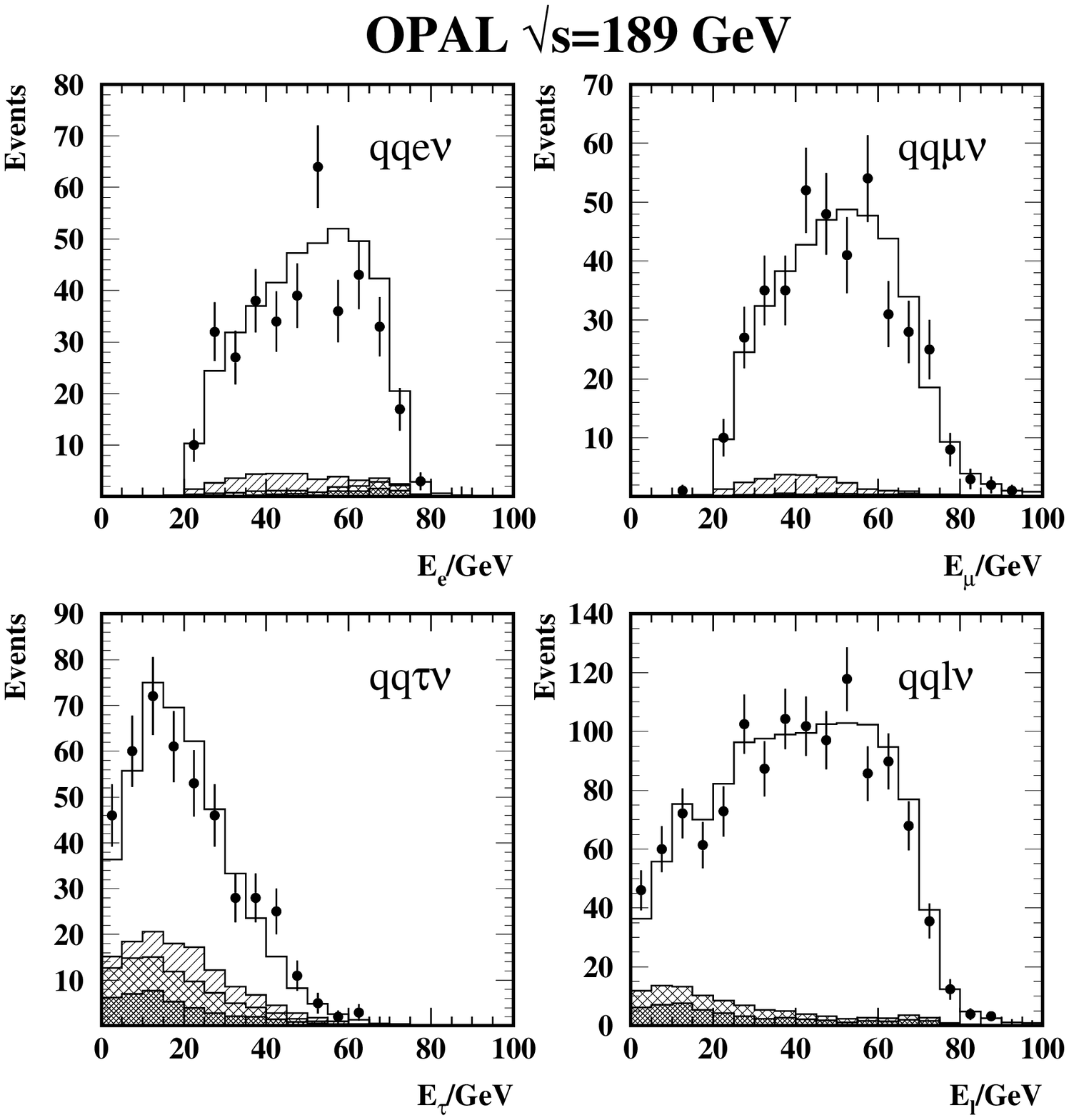,width=\textwidth}
   \caption{ 
     Distributions of measured lepton energy for events selected as
     \qqen, \qqmn, and \qqtn. 
     Also shown is the combined distribution for all events selected 
     as \WWqqln. 
     The data are shown as the points with statistical error bars, while
     the histogram is the total Monte Carlo expectation.
     The background from two-fermion processes is shown by the dark hatched
     region, while the non-\CC\ four-fermion background is shown by the
     lighter hatched region.
     Mis-identified \WWqqln\ events are shown in the individual lepton
     classifications as the singly hatched histogram.
     \label{fig:qqlnsel}
     }
 \end{center}
\end{figure}

The numbers of events selected in the individual \WWqqln\ lepton 
classes are summarized in Table~\ref{tab:events}, with a total
of 1246 events selected as inclusive \WWqqln\ candidates and 
$112 \pm 10$ expected from non-\CC\ background sources.
Figure~\ref{fig:qqlnsel} shows distributions of the reconstructed 
energy of the lepton in the \qqen, \qqmn, and \qqtn\ selection 
channels.
The data distributions are in good agreement with the Monte Carlo 
expectations.
%
%
\section{\boldmath \WWqqqq\ Event Selection}
\label{sec:qqqq}

The selection of fully hadronic \WWqqqq\ events is performed in two 
stages using a cut-based preselection followed by a likelihood
selection procedure similar to that used at 183~GeV~\cite{bib:opaltgc183}.
This likelihood selection is primarily designed to reject the dominant 
background from the $\eetoqq$ process where the di-quark system 
fragments into a four jet topology.
The changes from this previous selection are a different 
set of variables used for the preselection cuts and likelihood 
calculation, and a new method based on data for determining the 
accepted background rate.
No attempt is made to discriminate against the neutral current 
process \ZZqqqq.

All events which are classified as hadronic~\cite{bib:tkmh} and which 
have not been selected by either the \lnlns\ or the \qqlns\ selections 
are considered as candidates for the \WWqqqq\ selection.
In addition, any event which was rejected as a four-fermion background
event in the \qqlns\ selection is also rejected as a \qqqq\ candidate 
event.

Tracks and calorimeter clusters are combined into four jets using the
Durham algorithm~\cite{bib:durham} and the total momentum and energy of 
each jet is corrected for double-counting of energy~\cite{bib:GCE}.
To remove events which are clearly inconsistent with a fully hadronic 
\WW\ decay, candidate events are required to satisfy a set of 
preselection cuts including a cut on minimum visible energy (70\% of
\roots), minimum invariant mass (75\% of \roots), and minimum 
multiplicity per jet (one track).
The most important preselection cut is a limit on the logarithm of
the QCD matrix element for four jet production
$(\log_{10}(\WQCD) < 0)$~\cite{bib:qcd420}.
\WQCD\ is an event weight formed from the tree level ${\cal O}(\alpha_s^2)$ 
matrix element~\cite{bib:ERT} for the four jet production process 
($\eetoqq\rightarrow\qqqq,\qqgg$). 
The value of \WQCD\ is determined by using the observed 
momenta of the four reconstructed jets as estimates of the underlying
parton momenta which are input to the matrix element calculation.
The largest value of this matrix element calculated after
considering all 24 permutations of the jet-parton association in each event 
is found to have the best discriminating power between signal and background.

The preselection requirements reject an estimated $96$\% of the \eetoqq\ 
events which comprise the dominant source of background in the 
\WWqqqq\ event selection.
The preselection efficiency for the hadronic \WWqqqq\ decays 
is estimated to be $93$\%. 
A total of 2077 data events pass the preselection, of which 
775 are expected to be from non-\CC\ sources.

\begin{figure}[htbp]
  \begin{center}
    \epsfig{file=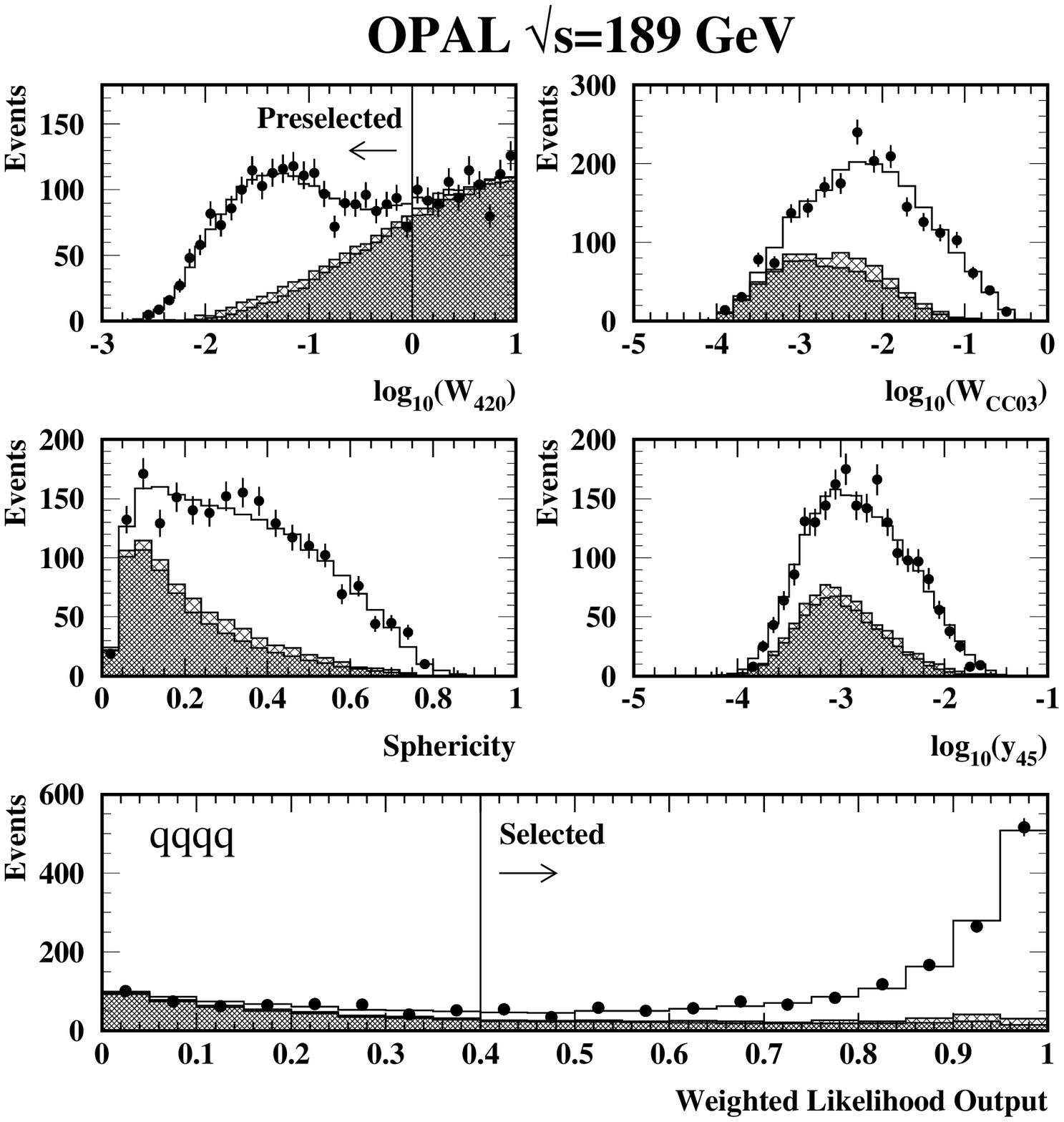,width=\textwidth}
    \caption{
      Distributions of the \WWqqqq\ likelihood selection variables.
      The four likelihood input variables and resulting
      likelihood discriminant are shown for all events
      passing the \WWqqqq\ preselection.
      The data are shown as the points with statistical error bars,
      while the histogram is the total Monte Carlo expectation.
      The dark hatched region shows the contribution from two-fermion
      processes, while the light hatched region shows the contribution from
      non-\CC\ four-fermion processes.
      In the \WQCD\ distribution, the events used to normalize
      the \eetoqq\ background rate are also shown to the right of the
      vertical line.
      In the likelihood output distribution, the selection cut used in the 
      counting analysis is indicated by the vertical line,
      with the selection accepting all events to the right of this cut.
      \label{fig:qqqqsel}
      }
  \end{center}
\end{figure}

Events satisfying the preselection cuts are classified as signal or
background based upon a four variable likelihood selection.
The following likelihood variables are selected to provide a good 
separation between the hadronic \WWqqqq\ signal and the \eetoqq\ four 
jet background, while minimizing the total number of variables used:
\begin{itemize}
        \item  $\log_{10}(\WQCD)$, the QCD four jet matrix element;
        \item  $\log_{10}(\WCC)$, the \EXCALIBUR\ matrix 
               element~\cite{bib:EXCALIBUR} for the \CC\ process (\WWqqqq);
        \item  $\log_{10}(y_{45})$, the logarithm of the value of the Durham 
               jet resolution parameter at which an event is reclassified from 
               four jets to five jets;
        \item  event sphericity.
\end{itemize}
Figure~\ref{fig:qqqqsel} shows the distribution of these four likelihood 
variables for all preselected events found in the 189~GeV data.
To improve the statistical power of this selection, a multi-dimensional
likelihood technique is used to account for the correlations between 
the four likelihood input variables~\cite{bib:PC}.
Most of the separation between the signal and background events is
provided by the two matrix element values $\log_{10}(\WCC)$ and
$\log_{10}(\WQCD)$, which give the relative probability that the 
kinematics of the observed event are consistent with signal or 
background production respectively.

An event is selected as a hadronic \WWqqqq\ candidate if the likelihood
discriminant variable, also shown in Figure
\ref{fig:qqqqsel}, is greater than 0.4.
This cut value was chosen to maximize the expected statistical power 
of this selection assuming the Standard Model rate for \CC\ production.
The efficiency of this likelihood selection for \WWqqqq\ events is 
estimated from \KORALW\ Monte Carlo samples to be $(86.4 \pm 0.9)\%$, 
where the error represents an estimate of the systematic uncertainties.
The individual components of this systematic uncertainty 
are shown in Table~\ref{tab:effsys}.

For the purposes of extracting a cross section, an alternative technique 
of weighting all preselected events according to the likelihood output 
is employed rather than selecting specific events by making a cut.
A similar method was used in previous results~\cite{bib:opalmw172},
although in this analysis the weights $(w_{i})$ are calculated for each
bin $(i)$ of the likelihood discriminant from the expected \CC\ signal
purity in that bin.
The cross section can then be expressed in terms of the weighted
values of efficiency, background, and observed events as
\begin{displaymath}
\sigma(\qqqq) = 
(\frac{1}{\cal L} \sum_i w_i N_i - \sum_i w_i \sbgd_i)/(\sum_i w_i \esig_i), 
\end{displaymath}
where ${\cal L}$ is the luminosity of the sample.
The values $N_i$, $\esig_i$, and $\sbgd_i$ are the observed events,
signal efficiency, and accepted background respectively in each bin.
The statistical uncertainty on the weighted number of events is given by
$\sqrt{\sum (w_i)^2 N_i}$, and by using this weighting technique an 
improvement of 3\% in the expected $\sigma(\qqqq)$ statistical error 
is gained.
Results for both techniques are presented in Section~\ref{sec:results}
and Tables~\ref{tab:back}--\ref{tab:effsys}.

The main systematic uncertainty on the selection efficiency
results from the modeling of the QCD hadronization process.
This uncertainty is estimated by comparing the selection
efficiency predicted using the \JETSET\ hadronization model with an
alternative model from the \HERWIG\ generator.
In addition, the effect of varying the parameters
$\sigma_{\mathrm{q}}$, $b$, $\Lambda_{\mathrm{QCD}}$, and $Q_{0}$ 
of the \JETSET\ hadronization model by one standard deviation 
about their tuned values~\cite{bib:qqqqQCD} is considered.
For these \JETSET\ tune studies, a fast parameterized simulation 
of the OPAL detector was used.
The Monte Carlo modeling of the \CC\ signal, including the 
detector simulation, is further studied by comparing
the distributions of the preselection and likelihood variables 
seen in data with various Monte Carlo estimates.
The signal efficiency determined by \KORALW\ is also compared to 
other generators (\EXCALIBUR, \PYTHIA, and \GRC) to test the Monte
Carlo description of the underlying hard process.
In each case, the observed differences are taken as an estimate of the
systematic uncertainty.
Possible biases related to final state interactions between the
hadronic systems produced by different W bosons have been evaluated
for color-reconnection effects~\cite{bib:CR} and Bose-Einstein 
correlations~\cite{bib:BE}.
These effects are found to be small, and the total change in 
predicted selection efficiency when these effects are included
in the hadronization model is taken as the systematic uncertainty.

The accepted \eetoqq\ background is estimated from \PYTHIA\ Monte Carlo 
samples, with \HERWIG\ and \KK\ being used as cross-checks.
All of these generators include only ${\cal O}(\alpha_s)$ matrix elements
for hard gluon emission, and rely upon a parton shower scheme
to predict the four jet production rate.
It has been suggested that this could lead to errors of up 
to 10\% in the rate of \eetoqq\ background when compared to a more 
complete ${\cal O}(\alpha_{s}^{2})$ matrix element 
approach~\cite{bib:Sterling}.
To reduce the uncertainty on this background estimate, a technique 
to measure this rate directly from the data has been used in this analysis.
By comparing the number of events seen in data and Monte Carlo 
in the range $(0<\log_{10}(\WQCD)<1)$ which would otherwise pass the
preselection cuts, the overall four jet rate predicted by the Monte
Carlo is normalized to the data.
A correction of $(-3.6 \pm 3.2)\%$ is found for the default
\PYTHIA\ sample assuming a total \eetoqq\ production cross section
of 98.4~pb, where the uncertainty is the statistical precision
of the normalization procedure.
The observed data and corrected Monte Carlo expectation in this 
`sideband' background region is shown in Figure~\ref{fig:qqqqsel}.
The expected contamination from \CC\ production in this region is 
less than 3\%, resulting in a negligible bias on the extracted \CC\ 
cross section.

Additional uncertainties on the background rate from the modeling
of the hadronization process are evaluated in the same manner as
the uncertainty on the signal efficiency.
The background normalization procedure has been consistently 
applied during these systematic checks.
Uncertainties in the non-\CC\ four-fermion background are 
estimated by comparing the expectations of \KORALW, \GRC, 
and \EXCALIBUR.
This background is predominantly from the neutral current process 
\ZZqqqq, of which only 15\% is in final states with direct 
interference with the \CC\ diagrams.
In each case, the single largest difference observed in a set of 
systematic checks is taken as an estimate of the uncertainty.

A total of 1546 \WWqqqq\ candidate events are selected
by the counting analysis, with an expected non-\CC\ background of
$325 \pm 21$ events.
Using the weighting technique, $1306 \pm 32$ weighted events
are observed with a weighted background estimate of $287 \pm 15$ 
events.
%
%
\section{\boldmath \WW\ Cross Section and W Decay Branching Fractions}
\label{sec:results}

The observed numbers of selected \WW\ events are used to measure the
\WW\ production cross section and the W decay branching fractions 
to leptons and hadrons.  
The measured cross section corresponds to that of W-pair production 
from the \CC\ diagrams as discussed earlier.
The expected four-fermion backgrounds quoted throughout this 
paper include contributions from both non-\CC\ final states and 
the effects of interference with the \CC\ diagrams.  
Mis-identified \CC\ final states are not included in the background
values listed in Table~\ref{tab:back}, but rather are
taken into account by off-diagonal entries in the efficiency matrix
shown in Table~\ref{tab:eff}.
 
Table~\ref{tab:events} summarizes the event selections in the
three \WW\ decay topologies. 
The expected numbers of events assume 
a center-of-mass energy of $\rroots\pm\errroots$~GeV, 
an integrated luminosity of $\intLdtfull\pm\dLtot\ \mathrm{pb}^{-1}$, 
and a \WW\ cross section of $\SMDPAxs\pm\dSMDPAxs$~pb as predicted by 
the calculations of \YFSWW\ and \RACWW.

As in~\cite{bib:opaltgc183}, the \WW\ cross section and branching 
fractions are measured using data from the ten separate decay channels. 
Three different fits are performed with all correlated systematic
uncertainties taken into account.
In the first case
\sigccthree(189~GeV), $\Br(\Wtoen)$, $\Br(\Wtomn)$, and $\Br(\Wtotn)$ are
extracted under the assumption that
\begin{eqnarray*}
     \Br(\Wtoen)+\Br(\Wtomn)+\Br(\Wtotn)+\Br(\Wtoqq) & = & 1.
\end{eqnarray*}
In the second fit, the additional constraint of charged current lepton 
universality is imposed. 
The results of these branching fraction fits to the 189~GeV data alone
are summarized in Table~\ref{tab:xsecbr_results} along with the
Standard Model expectation, which is estimated to have a theoretical
uncertainty of 0.1\%~\cite{bib:LEP2YR}.

From this second fit, the \WW\ \CC\ production cross sections in each
channel can be extracted under the assumption of lepton universality, 
assuming Standard Model rates for all other processes:
\begin{eqnarray*}
 \sigma(\WWlnln) & = & 1.64 \pm 0.11 \stat \pm 0.03 \syst \, {\mathrm{pb}}, \\
 \sigma(\WWqqln) & = & 7.04 \pm 0.22 \stat \pm 0.10 \syst \, {\mathrm{pb}}, \\
 \sigma(\WWqqqq) & = & 7.68 \pm 0.24 \stat \pm 0.16 \syst \, {\mathrm{pb}}.
\end{eqnarray*}
These results are consistent with the Standard Model expectations 
of 1.72~pb, 7.13~pb, and 7.41~pb respectively.
The cross section in the \qqqqs\ channel has been determined using
the event weight technique described in Section~\ref{sec:qqqq}.
Using the counting method yields a consistent result of
$\sigma(\WWqqqq) =  7.70 \pm 0.25 \pm 0.18 \ {\mathrm{pb}}$.

In the third fit, all W decay branching fractions are fixed to the values
predicted by the Standard Model, and the \WW\ cross section is determined 
to be 
\begin{eqnarray*}
 \sigccthree(189~{\mathrm{GeV}})=\xsecresult~{\mathrm{pb}},
\end{eqnarray*}
consistent with the Standard Model expectation of $\SMDPAxs\pm\dSMDPAxs$~pb. 
%
%
\section{\boldmath $\epem\rightarrow\lplm\nunu$ Cross Section Measurement}
\label{sec:lnln4f}

The fully leptonic event selection has only a small
($38 \pm 10$~fb) contamination of background expected from sources 
without two leptons and two neutrinos in the final state.
It is therefore well suited to measuring the inclusive
four-fermion cross-sections for the six charged 
di-lepton final states which within the Standard Model
receive contributions from some or all of the WW, ZZ, \Wenu, \Zee,
and \Znunu\ diagrams and in particular their respective interferences.

The four-fermion $\epem\rightarrow\lplm\nunu$ cross sections are defined 
in terms of the following kinematic acceptance cuts:
\begin{itemize}
\item 
  at least one of the charged leptons is produced with 
  $|\cos\theta|<0.90$, where the angle $\theta$ is the scattering
  angle between the outgoing lepton and the incoming electron;
\item
  both charged leptons are produced with $|\cos\theta|<0.99$;
\item
  the invariant mass calculated from the four-momentum balancing the
  two prompt neutrinos must be greater than 10~GeV, while the transverse
  momentum of this recoil four-momentum must have
  $\Pt/\Ebeam > 5\%$.\footnote{This definition of the `visible' system
    in terms of the recoil from the two neutrinos avoids ambiguities 
    in defining the invariant mass when there is additional photon radiation.}
\end{itemize}
For this theoretical signal definition, $\tau$ leptons are treated
as stable such that the prompt $\tau$ lepton is used in the kinematic
acceptance cuts.
This signal definition is chosen to approximate the kinematic acceptance of
the \lnlns\ selection.
The efficiency of the inclusive $\lnlns$ selection for this
signal definition is 86\%.

A combined fit for the six four-fermion cross sections is performed
and summarized in Table~\ref{tab:xsec-lnln}.
The observed cross sections are in good agreement with the
Standard Model rates predicted by the \KORALW\ four-fermion generator.
%
%
\section{Combination with Previous Data}
\label{sec:comb}

A simultaneous fit to the numbers of \WW\ candidate events in the 
ten identified final states (\enens, \mnmns, \tntns, \enmns, \entns, 
\mntns, \qqen, \qqmn, \qqtn, and \qqqq) observed by OPAL at 
center-of-mass energies of 161~GeV, 172~GeV, 183~GeV, and 189~GeV 
gives the following values for the leptonic branching fractions 
of the W boson:
\begin{eqnarray*}
  \Br(\Wtoen) & = &  10.46 \pm 0.42 \stat \pm 0.14 \syst \, \% \\ 
  \Br(\Wtomn) & = &  10.50 \pm 0.41 \stat \pm 0.12 \syst \, \% \\ 
  \Br(\Wtotn) & = &  10.75 \pm 0.52 \stat \pm 0.21 \syst \, \%.
\end{eqnarray*}
Correlations between the systematic uncertainties at the different
energy points have been accounted for in the fit.
These results are consistent with the hypothesis of lepton universality, 
and agree well with the Standard Model prediction of \SMbrlv\%.
The correlation coefficient for the resulting values of 
$\Br(\Wtoen)$ and $\Br(\Wtomn)$ is $-0.05$. 
The correlation coefficient for the results of either 
$\Br(\Wtoen)$ or $\Br(\Wtomn)$ with the measurement of 
$\Br(\Wtotn)$ is $-0.25$.

A simultaneous fit assuming lepton universality gives
\begin{eqnarray*} 
  \Br(\Wtoqq) & = & \brqqresult,
\end{eqnarray*}
which is consistent with the Standard Model expectation of \SMbrqq\%.
Here, the dominant sources of systematic uncertainty are from the
uncertainty on the \eetoqq\ background in the \WWqqqq\ channel
and the uncertainties on the \WWqqln\ and \WWqqqq\ selection 
efficiencies.

The hadronic branching fraction can be interpreted as a measurement of
the sum of the squares of the six elements of the CKM
mixing matrix, \Vij, which do not involve the top quark:
\begin{eqnarray*}
     {{\Br(\Wtoqq)}\over{(1-\Br(\Wtoqq))}} & = &
 \left( 1+\frac{\alpha_s(\Mw)}{\pi} \right)
\sum_{i={\mathrm{u,c}}; \, j={\mathrm{d,s,b}}} \Vij^2.
\end{eqnarray*}
The theoretical uncertainty of this improved Born approximation 
due to missing higher order corrections is estimated to 
be 0.1\%~\cite{bib:LEP2YR}.
Taking $\alpha_s(\Mw)$ to be $0.120\pm0.005$,
the branching fraction $\Br(\Wtoqq)$ from the 161 -- 189~GeV
data yields
\begin{eqnarray*}
 \sum_{i={\mathrm{u,c}};\, j={\mathrm{d,s,b}}}
  \Vij^2 & = & 2.077\pm0.059\mathrm{(stat.)}\pm0.027\mathrm{(syst.)},
\end{eqnarray*}
which is consistent with the value of 2 expected from unitarity in
a three-generation CKM matrix.

Using the experimental knowledge of the sum,
$\Vud^2+\Vus^2+\Vub^2+\Vcd^2+\Vcb^2 = 1.048\pm0.007$~\cite{bib:pdg}, 
the above result can be interpreted as a measure of \Vcs\ which is the 
least well determined of these matrix elements:
\begin{eqnarray*}
  \Vcs & = & 1.015\pm0.029\mathrm{(stat.)}\pm0.013\mathrm{(syst.)}.
\end{eqnarray*}
The uncertainty in the sum of the other five CKM matrix elements,
which is dominated by the uncertainty on $\Vcd$, contributes
a negligible uncertainty of 0.004 to this determination of $\Vcs$.
A more direct determination of $\Vcs$ is also performed by OPAL in 
the measurement of the hadronic branching fraction of the W boson
to charm quarks~\cite{bib:opalrcw}.
%
%
\section{Summary}

\begin{figure}[tbhp]
  \begin{center}
    \epsfig{file=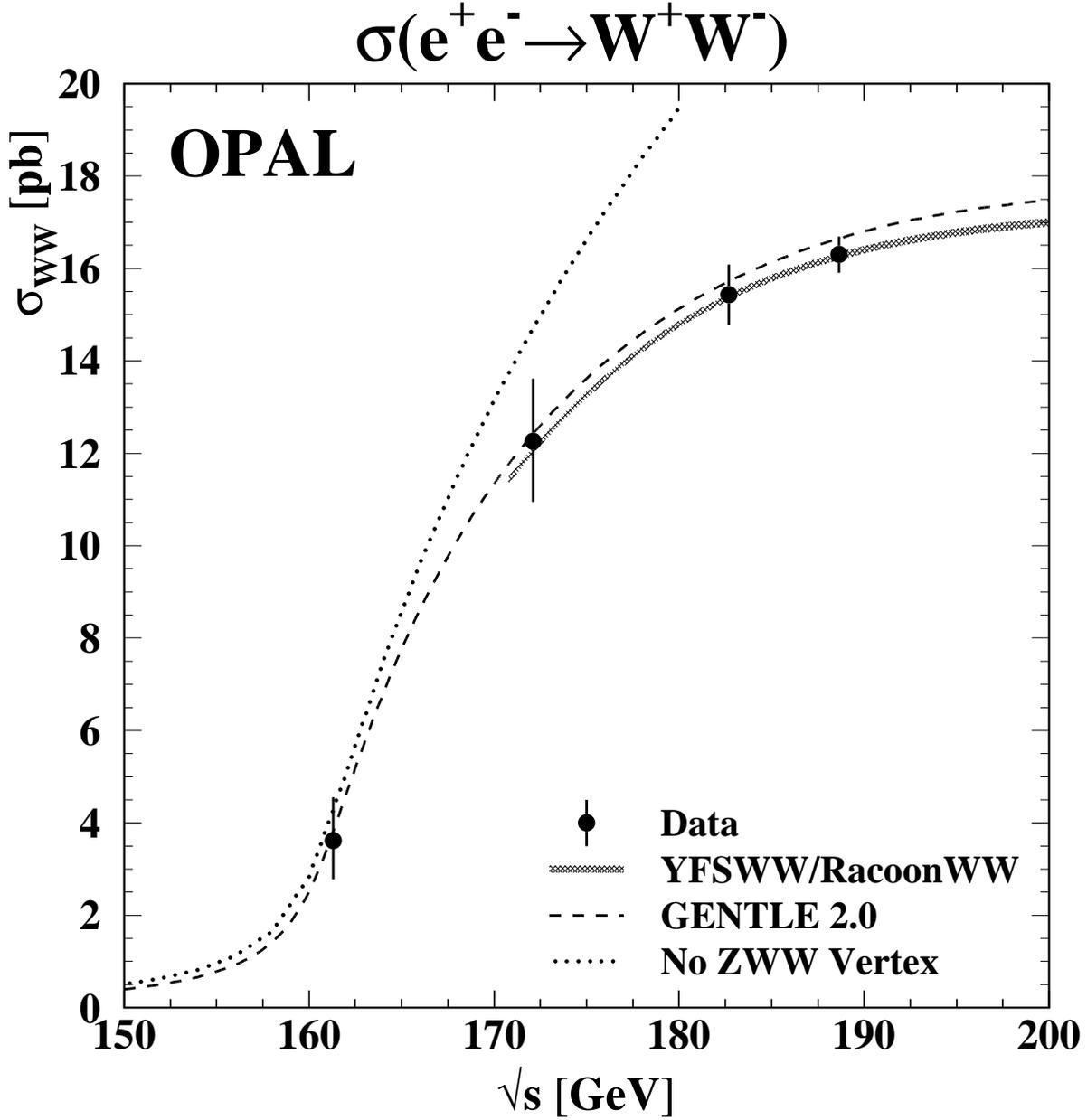,width=\textwidth}
    \caption{
      The dependence of \sigccthree\ on $\protect\roots$.
      The \WW\ cross sections measured at 
      $\protect\roots = 188.6$~GeV (this paper), at 
      $\protect\roots = 161.3$~GeV\protect\cite{bib:opalmw161}, at
      $\protect\roots = 172.1$~GeV\protect\cite{bib:opalmw172}, and at 
      $\protect\roots = 182.7$~GeV\protect\cite{bib:opaltgc183} are
      shown. 
      The error bars include statistical and systematic contributions. 
      The shaded area shows the prediction
      of \RACWW\ and \YFSWW, with the width of the band covering the
      range of the estimated theoretical uncertainty.
      The dashed curve shows the older \GENTLE~2.0 prediction,
      while the dotted curve indicates the expected cross section if there
      is no ZWW coupling.
      \label{fig:sigmaww} 
      }
  \end{center}
\end{figure}

Using \intLdt~pb$^{-1}$ of data recorded by OPAL at a mean 
center-of-mass energy of $\roots = 188.6$~GeV, a total of 
\nsel\ W-pair candidate events are selected.
The data are used to determine the \CC\ production cross section
assuming Standard Model decay rates:
\begin{eqnarray*}
  \sigccthree(189~{\mathrm{GeV}}) & = & \xsecresult \, \mathrm{pb}.
\end{eqnarray*}
The measured \WW\ production cross section at $\roots = 188.6$~GeV is
shown in Figure~\ref{fig:sigmaww}, together with the previous OPAL
measurements of \sigccthree\ at $\roots = 161.3$~GeV~\cite{bib:opalmw161},
$\roots = 172.1$~GeV~\cite{bib:opalmw172}, and at 
$\roots = 182.7$~GeV~\cite{bib:opaltgc183}.
The measured cross sections clearly favor the Standard Model prediction 
over the model where there is no coupling between the weak gauge bosons, 
confirming the non-Abelian nature of the electroweak interaction.
When combined with previous OPAL measurements under the assumption 
of lepton universality, the hadronic branching fraction of the 
W boson is found to be
\begin{eqnarray*}
  \Br(\Wtoqq) & = \brqqresult,
\end{eqnarray*}
which is consistent with the Standard Model expectation of 67.5\%.

Similar measurements have been made at $\roots \leq 189$~GeV
by ALEPH~\cite{bib:aleph}, DELPHI~\cite{bib:delphi}, 
and L3~\cite{bib:l3}.
Results consistent with the Standard Model are observed by all 
four LEP collaborations.
%
%
\section*{Acknowledgements}
We particularly wish to thank the SL Division for the efficient 
operation of the LEP accelerator at all energies
and for their continuing close cooperation with
our experimental group.  
We thank our colleagues from CEA, DAPNIA/SPP, CE-Saclay for their 
efforts over the years on the time-of-flight and trigger
systems which we continue to use.  
In addition to the support staff at our own
institutions we are pleased to acknowledge the \\
Department of Energy, USA, \\
National Science Foundation, USA, \\
Particle Physics and Astronomy Research Council, UK, \\
Natural Sciences and Engineering Research Council, Canada, \\
Israel Science Foundation, administered by the Israel
Academy of Science and Humanities, \\
Minerva Gesellschaft, \\
Benoziyo Center for High Energy Physics,\\
Japanese Ministry of Education, Science and Culture (the
Monbusho) and a grant under the Monbusho International
Science Research Program,\\
Japanese Society for the Promotion of Science (JSPS),\\
German Israeli Bi-national Science Foundation (GIF), \\
Bundesministerium f\"ur Bildung und Forschung, Germany, \\
National Research Council of Canada, \\
Research Corporation, USA,\\
Hungarian Foundation for Scientific Research, OTKA T-029328, 
T023793 and OTKA F-023259.\\
%
%

%
%
\pagebreak
\renewcommand{\arraystretch}{1.1}
\begin{table}
\centering
\begin{tabular}{l|cccccc|ccc|c}
\hline\hline
Event & 
\multicolumn{10}{c}{Efficiencies (\%) for $\WW\rightarrow$} \\
Selection &
\senen$\!\!$&\smnmn$\!\!$&\stntn$\!\!$&
\senmn$\!\!$&\sentn$\!\!$&\smntn$\!\!$&
\qqen$\!\!\!$&\qqmn$\!\!\!$&\qqtn$\!\!\!$&\qqqq$\!\!\!\!$\\
\hline
\senen & 75.5 &  0.0 &  1.0 &  0.1 &  6.2 &  0.0 &  0.0 &  0.0 &  0.0 &  0.0 \\
\smnmn &  0.0 & 80.4 &  0.6 &  1.2 &  0.1 &  6.1 &  0.0 &  0.0 &  0.0 &  0.0 \\
\stntn &  0.5 &  0.3 & 46.4 &  0.4 &  4.1 &  5.0 &  0.0 &  0.0 &  0.0 &  0.0 \\
\senmn &  2.5 &  0.4 &  1.2 & 77.8 &  6.2 &  7.2 &  0.0 &  0.0 &  0.0 &  0.0 \\
\sentn &  8.5 &  0.0 & 11.1 &  3.9 & 63.0 &  1.1 &  0.0 &  0.0 &  0.0 &  0.0 \\
\smntn &  0.1 &  6.6 &  8.3 &  3.9 &  0.8 & 60.6 &  0.0 &  0.0 &  0.0 &  0.0 \\
\hline
\qqen &   0.0 &  0.0 &  0.1 &  0.0 &  0.2 &  0.0 & 85.4 &  0.1 &  3.8 &  0.0 \\
\qqmn &   0.0 &  0.0 &  0.0 &  0.0 &  0.0 &  0.2 &  0.1 & 89.2 &  4.3 &  0.1 \\
\qqtn &   0.0 &  0.0 &  0.4 &  0.0 &  0.0 &  0.0 &  4.5 &  4.4 & 68.4 &  0.8 \\
\hline
\qqqq &   0.0 &  0.0 &  0.0 &  0.0 &  0.0 &  0.0 &  0.0 &  0.1 &  0.6 & 86.4 \\
Weighted& 0.0 &  0.0 &  0.0 &  0.0 &  0.0 &  0.0 &  0.0 &  0.1 &  0.6 & 72.3 \\
\hline\hline
\end{tabular}
\caption{
  \CC\ selection efficiency matrix.
  For the \WWqqqq\ selection the efficiencies are listed for both
  the counting and weighted event selections as described in the text.
}
\label{tab:eff}
\end{table}

\begin{table}[tb]
\centering
\begin{tabular}{l|cccccc|c}
\hline\hline
  Background & \multicolumn{7}{c}{Accepted background cross sections (fb)} \\
  Class & \epem & \mpmm & \tptm & \epmmmp & \epmtmp & \mpmtmp & \lplm \\
\hline
Non-\lplm\nunu    &  0.7 &  1.1 & 19.8 &  1.9 &  9.6 &  4.4 &  37.5 \\
\llnn             &  6.6 &  9.1 &  8.6 &  4.5 &  9.9 &  6.5 &  45.2 \\
\lnln (4f - \CC)  & 10.9 & 10.0 & 13.9 &  9.5 & 21.5 & 11.6 &  77.4 \\
\hline
Total Background  & 18.2 & 20.2 & 42.3 & 15.9 & 41.0 & 22.5 & 160.1 \\
\hline\hline
\end{tabular}
\caption{
  Accepted \lnlns\ background cross sections listed by selection class.
  Non-\lplm\nunu\ is defined as final states which do not contain
  two leptons and two neutrinos, while \llnn\ are final states only
  produced by neutral current processes.
  The (4f - \CC) background is the difference in accepted \lnlns\
  cross section between complete four-fermion production and
  \CC-only production for \lnln\ final states.
}
\label{tab:backlnln}
\end{table}

\newcommand{\dash}{\multicolumn{2}{c}{--}}

\begin{table}[tb]
\centering
\begin{tabular}{l|r@{$\ \pm \ $}lr@{$\ \pm \ $}l
    r@{$\ \pm \ $}lr@{$\ \pm \ $}lr@{$\ \pm \ $}l
    r@{$\ \pm \ $}l} 
\hline\hline
  \multicolumn{1}{l|}{} &
  \multicolumn{12}{c}{Accepted background cross sections (fb)} \\
  \multicolumn{1}{l|}{} &
  \multicolumn{12}{c}{Event Selection $\WW\rightarrow$ } \\
  \multicolumn{1}{l|}{Source} & 
  \multicolumn{2}{c}{\lnln} & 
  \multicolumn{2}{c}{\qqen} & 
  \multicolumn{2}{c}{\qqmn} & 
  \multicolumn{2}{c}{\qqtn} &
  \multicolumn{2}{c}{\qqqq} &
  \multicolumn{2}{c}{Weighted} \\ 
\hline 
  \lnln     & 77&10 &  \dash  &\dash  &   0&1  &  \dash   &  \dash  \\
  \qqln     &  3&1  &  54&21  &  2&1  &  74&8  &  \dash   &  \dash  \\
  \qqqq     &\dash  &   0&1   &  1&1  &  13&3  &  392&67  &  320&45 \\
\hline
  \llnn     & 45&2  & \dash   &\dash  & \dash  &  \dash   &  \dash  \\
  \qqnn     &  3&1  & \dash   &\dash  &  32&3  &  \dash   &  \dash  \\
  \qqee     &  1&1  &  30&7   &\dash  &  48&12 &   22&6   &   19&6  \\
  \qqll     &\dash  &   2&1   & 28&2  &  46&4  &   26&5   &   22&4  \\
  \eeff     &  9&9  &   7&7   &  1&1  &   7&7  &    2&2   &    2&2  \\ 
\hline
  \nngg     &  2&1  & \dash   &\dash  & \dash  &  \dash   &  \dash  \\
  \lplm     & 20&4  &   2&1   &  1&1  &   5&1  &  \dash   &  \dash  \\
  \qq       &\dash  &  40&8   & 22&4  & 192&29 & 1337&87  & 1210&62 \\
\hline
  Combined  &160&14 & 135&24   & 55&6  & 417&34 & 1777&117 & 1570&84 \\ 
\hline\hline
\end{tabular}
\caption{
  Accepted background cross sections.
  Backgrounds from non-\CC\ sources are shown for the 189~GeV 
  \WW\ selections.
  The first three lines list differences in accepted cross sections
  between a complete four-fermion sample and a \CC\ sample for
  these final states.
  The \eeff\ class, containing all additional four-fermion
  background not explicitly counted elsewhere, is dominated
  by two-photon interactions.
  Backgrounds in the \qqqqs\ selection are listed for both the 
  counting and weighted event analyses as described in the text.
  All errors include both statistical and systematic contributions.
}
\label{tab:back}
\end{table}

\begin{table}[htbp]
 \begin{center}
 \begin{tabular}{l|cr@{$\ \pm \ $}lr@{$\ \pm \ $}lr@{$\ \pm \ $}l} 
\hline\hline
  Selected as & Observed & 
  \multicolumn{2}{c}{Total Expected} & 
  \multicolumn{2}{c}{Efficiency (\%)} &
  \multicolumn{2}{c}{Background} \\
\hline
  \enen &   37 &   34.4 & 1.1  & 75.5 & 1.1 &   3.3 &  1.0 \\
  \mnmn &   34 &   37.1 & 1.1  & 80.4 & 1.2 &   3.7 &  1.0 \\
  \tntn &   37 &   30.9 & 1.0  & 46.4 & 0.7 &   7.7 &  1.0 \\
  \enmn &   68 &   68.0 & 1.3  & 77.8 & 1.2 &   2.9 &  1.0 \\
  \entn &   46 &   61.9 & 1.2  & 63.0 & 1.0 &   7.5 &  1.0 \\
  \mntn &   54 &   55.0 & 1.2  & 60.6 & 0.9 &   4.1 &  1.0 \\
\hline
  \qqen &  389 &  414.3 &  6.5 & 85.4 & 0.8 &  25.1 &  4.4 \\
  \qqmn &  420 &  418.9 &  4.8 & 89.2 & 0.8 &  10.1 &  1.1 \\
  \qqtn &  437 &  423.6 &  9.5 & 68.4 & 1.4 &  76.3 &  6.2 \\
\hline
  \lnln &  276 &  287.2 &  5.0 & 82.1 & 1.2 &  29.3 &  2.9 \\
  \qqln & 1246 & 1256.8 & 16.1 & 86.8 & 0.9 & 111.5 &  9.5 \\
  \qqqq & 1546 & 1500.0 & 25.2 & 86.4 & 0.9 & 325.3 & 21.4 \\
  Weighted & $1306.1 \pm 31.8$ & 1270.5 & 19.1 & 72.3  & 0.7 & 287.4 & 15.4 \\
\hline 
  any   & 3068 & 3044.0 & 32.1 & 86.6 & 0.6 & 466.1 & 23.6 \\
\hline\hline
\end{tabular}
\end{center}
\caption{
  Event selection summary.
  The observed and expected numbers of events for each
  selection category are shown for an integrated luminosity of
  $\intLdtfull\pm\dLtot$~pb$^{-1}$ at 
  $\roots = \rroots\pm\errroots$~GeV
  assuming Standard Model production rates.
  The expected efficiency 
  of each individual selection for that particular \CC\ final state
  and the expected number of background events from non-\CC\ processes
  is also shown separately.  
  The errors on the expected numbers of events includes the 
  theoretical uncertainty on the predicted \WW\ production 
  cross section of $\SMDPAxs\pm\dSMDPAxs$~pb.
  The inclusive $\WW\rightarrow\mathrm{any}$ numbers use the counting
  method in the \qqqq\ channel.
  The uncertainties on combined numbers account for all correlations.
}
\label{tab:events}
\end{table}

\begin{table}[htb]
\centering
\begin{tabular}{l|ccc|cc} 
  \hline\hline
  \multicolumn{1}{l|}{} &
  \multicolumn{5}{c}{Signal efficiency  error (\%) }    \\
  \multicolumn{1}{l|}{} &
  \multicolumn{5}{c}{Event Selection $\WW\rightarrow$ } \\
  Source of uncertainty & \qqen & \qqmn & \qqtn  & \qqqq & Weighted\\ 
\hline
Statistical                     & 0.18  & 0.17 & 0.27 & 0.06 & 0.20 \\     
Comparison of MC generators     & 0.24  & 0.51 & 0.78 & 0.21 & 0.23 \\ 
Fragmentation                   & 0.40  & 0.40 & 0.60 & 0.72 & 0.54 \\
\Mw\ dependence                 & 0.01  & 0.02 & 0.02 & 0.05 & 0.05 \\ 
Beam energy dependence          & 0.01  & 0.02 & 0.01 & 0.01 & 0.01 \\
Tracking losses                 & 0.40  & 0.10 & 0.30 & 0.18 & 0.15 \\
Final-state interactions        &  --   &  --  &  --  & 0.36 & 0.34 \\
Data/MC preselection            & 0.29  & 0.24 & 0.58 &  --  &  --  \\ 
Data/MC likelihood selection    & 0.36  & 0.34 & 0.72 & 0.12 & 0.17 \\ 
Detector occupancy              & 0.02  & 0.02 & 0.02 &  --  &  --  \\
Four-fermion rejection          & 0.30  & 0.10 & 0.30 &  --  &  --  \\ 
\hline
Total                           & 0.84  & 0.80 & 1.44 & 0.86 & 0.74 \\ 
\hline\hline
\end{tabular}
\caption{ 
   Selection efficiency systematic uncertainties.
   For the \WWqqqq\ selection the uncertainties are listed 
   for both the counting and weighted event selections.
   All contributions are listed as the absolute difference
   in selection efficiency.
}
 \label{tab:effsys}
\end{table}

\begin{table}[htbp]
 \centering
 \begin{tabular}{l|cc|c} 
    \hline\hline
Fitted    & \multicolumn{2}{c|}{Fit assumptions: } & Standard Model \\   
Parameter & No lepton universality & Lepton universality & Expectation \\ 
\hline
$\Br(\Wtoen)$ & $10.03\pm0.47\pm0.16$ \% &                          & 
\SMbrlv \% \\ 
$\Br(\Wtomn)$ & $10.63\pm0.47\pm0.16$ \% & $10.51\pm0.23\pm0.12$ \% & 
\SMbrlv \% \\ 
$\Br(\Wtotn)$ & $10.94\pm0.59\pm0.25$ \% &                          & 
\SMbrlv \% \\ 
\hline
$\Br(\Wtoqq)$ & $68.41\pm0.71\pm0.36$ \% & $68.47\pm0.70\pm0.35$ \% & 
\SMbrqq \% \\ 
\hline\hline
\end{tabular}
\caption{ 
   Summary of branching fraction results from the OPAL $\roots = 188.6$~GeV 
   data alone. 
   The results from the different branching fraction fits described in the 
   text are shown, where the two errors represent the statistical 
   and systematic uncertainties respectively.
   The leptonic and hadronic branching fractions in each case are not 
   independent, but have been determined assuming that the sum is 
   equal to unity.
   When assuming lepton universality, small differences in the 
   leptonic branching fractions due to lepton mass effects have 
   been neglected.
   }
\label{tab:xsecbr_results}
\end{table}

\begin{table}[htbp]
\centering
\begin{tabular}{l|cc}
   \hline\hline
   $\epem\rightarrow$ & Measured cross section (fb) & Expected (fb) \\
   \hline
   $\epem\nunu$                        & $290^{+56}_{-51}\pm 11 $ &  262 \\
   $\mpmm\nunu$                        & $195^{+44}_{-40}\pm 08 $ &  221 \\
   $\tptm\nunu$                        & $290^{+73}_{-66}\pm 13 $ &  207 \\
   $\mathrm{e}^{\pm} \mu^{\mp} \nunu$  & $384^{+60}_{-55}\pm 11 $ &  387 \\
   $\mathrm{e}^{\pm} \tau^{\mp} \nunu$ & $225^{+64}_{-58}\pm 10 $ &  388 \\
   $\mu^{\pm} \tau^{\mp} \nunu$        & $348^{+69}_{-63}\pm 11 $ &  376 \\
   \hline\hline
\end{tabular}
\caption{
  Four-fermion \lplm\nunu\ cross sections.
  Observed cross sections are shown for each \lnlns\ decay topology
  using the four-fermion signal definition described in the text.
  All six cross sections were determined in a simultaneous fit to
  the observed number of events.
  The errors shown are the statistical and systematic contributions
  respectively.
  The expected cross sections within the Standard Model are calculated
  using the \KORALW\ four-fermion Monte Carlo generator.
}
\label{tab:xsec-lnln}
\end{table}
%
%
\clearpage

\end{document}